\documentclass[tc,final]{copernicus}
\usepackage{graphicx}  
\usepackage{dcolumn}   
\usepackage{bm}        
\usepackage{amssymb}   
\usepackage{natbib}
\begin{document}
\nolinenumbers



\title{How far can we trust climate change predictions?}

\author{Francois~Louchet}
\affil{Professor Emeritus at Grenoble University (Grenoble Institute of Technology), formerly at "Laboratoire de Glaciologie et de Géophysique de l'Environnement", France}
\correspondence{F. Louchet (f.louchet@gmail.com)}

\runningtitle{How far can we trust climate change predictions?}
\runningauthor{Louchet}

\firstpage{1}

\maketitle

\begin{abstract}
Current techniques for predicting climate change are mainly based on "massive" deterministic numerical modeling. However, the ocean-atmosphere system is a so-called "complex system", made up of a large number of interacting elements. We show that, in such systems, owing to the particularly large sensitivity to initial conditions, the approach of a possible tipping over a critical point cannot be evidenced "by construction" using numerical modeling, due to the divergence of computation time in the vicinity of the tipping point.
On the other hand, the increasing amplitudes of observed climatic instabilities seem to be an obvious sign of the approach of such a tipping point, easily interpreted as a "critical softening", well known in the theory of dynamical systems, that would bring us irreversibly into a new and totally unexplored equilibrium state, except for a significantly higher temperature and in a much closer time than expected from numerical modeling extrapolations.
Thus, maintaining climate warming around 1.5$^o$C or 2$^o$C by 2030 or 2050 appears fairly unrealistic unless worldwide drastic green house gases reduction measures are immediately taken and applied.

\end{abstract}

\introduction

The Earth is currently undergoing an unprecedented climate change on a human scale, up to a possible destabilization, owing to the suddenness (on a geophysical scale) of the artificial injection of greenhouse gases into the atmosphere, resulting in a significant temperature increase. 
In such a context, a key point would be to understand and predict what the next equilibrium will look like, and what would be the levels of the associated climate parameters (temperature, humidity, precipitation, new patterns of atmospheric circulation and ocean currents, etc.). 

There are currently two different, and to some extent complementary, approaches: deterministic numerical modeling, which can be referred to as a "direct method", and a non-deterministic macroscopic analysis based on the theories of critical phenomena and of dynamical systems, founded by Henri Poincaré \citep{holmes1990poincare}, which can be called an "inverse method". These two methods lead to very different results for long-term climate predictions, the former predicting a continuous increase of temperatures, and the latter an irreversible tipping (critical point) beyond which a new and significantly warmer equilibrium state would be found.

\section{The direct method (numerical modeling)}
Most current climate prediction techniques are based on "massive" numerical simulations of the evolution of the "ocean-atmosphere" system. This system is subdivided into elementary cells, analogous to the pixels of digital images, but in 3 dimensions (and even in 4 dimensions if we include time). To put it simply, each of these cells is assigned an initial state, including values for temperature, humidity, received and emitted radiations, atmospheric pressure, wind or sea currents speeds and directions, greenhouse gas (GHG) content, etc... This is supplemented with possible evolution of GHG injection, that depends on more or less drastic political scenarios. The laws of physics (essentially the fundamental law of dynamics and the law of mass and energy conservation) are then applied to all these input data, completed by empirical laws parametrized by fitting on well-documented previous episodes \citep{malardel2009fondamentaux}
, and by the positive and negative feedback processes that are already known. The calculation is then launched. Each cell evolves according to its environment, "exchanging data" with its neighbours.

It turns out that in all cases the simulated average temperatures increase continuously with the amount of injected GHGs, more or less rapidly depending on the more or less voluntarist character of the emission control. But it is becoming increasingly clear that such forecasts, from one IPCC report to the next, must be constantly revised upwards. This is usually not "explained", but simply described by an increase in the so-called "temperature sensitivity to a doubling of the atmospheric $\rm{CO_2}$ content (...) \underline{whose origin remains mysterious} in spite of the numerous works devoted to it" \citep{Sherwood&al2020,SaintMartin&al2021}, suggesting at least a significant non-linearity of the system response, if not more.

Indeed, it is well known that, as a rubber band is continuously stretched, after deforming reversibly for a while, it eventually breaks. In the same way, if we let more and more snow accumulate on a shed roof, after a progressive bending of the frame, the structure will finally collapse. Continuously evolving causes may sometimes have discontinuous consequences. It is therefore justified to wonder whether, in the case of climate, we are really heading towards a dramatic but continuous warming, or instead towards a speeding up leading to a rapid and irreversible tipping into a still unknown state. 

In order to address this question in a relevant way, the most obvious idea should be to increase the accuracy of numerical calculations. In photography, it is well known that increasing the number of sensor pixels improves the "resolution" (i.e., the sharpness) of an image. By analogy, it may seem necessary to improve the accuracy and reliability of numerical calculations by reducing the cell size, and therefore by increasing their number in proportion. But, in contrast with a camera sensor, we are dealing here with a so-called "complex system", made up of a very large number of interacting elements. The analysis of such systems requires taking into account the behaviour of each of these elements (similar to brain neurons), but also (and more especially) of all their interactions (conveyed by synapses). 

One of the characteristics (and beauty) of complex systems is the emergence of a global "macroscopic" behaviour which is not present at the level of basic elements. This is the case for sheep herds, fish schools, or crowds, whose global properties cannot be considered as the simple addition of individual ones. Extrapolating the behaviour of an isolated sheep cannot help predict that of the herd. 

A fundamental property of complex systems is that the slightest inaccuracy in the input data may have significant (not to say major) consequences on the overall behaviour. This is the well known "\underline{butterfly wing effect}". Nevertheless, in order to improve the forecasts "resolution", and thus presumably their reliability, numerical modeling usually reduces the cell size as much as possible, which necessarily increases their number. Unfortunately, the number of their mutual interactions increases significantly faster. As a result, beyond a given level, the number of elements and interactions that the numerical model has to take into account becomes readily out of control. We then have the choice between either increasing the number of machines constituting the network, or reducing the number of elements to be processed. In this last case, such a reduction is generally carried out using various techniques that in fact amount to increase the cell size. Whatever they are, they bring us back to the "starting point", replacing a detailed information by a coarser one. The picture that was starting to be sharpened becomes blurred again.

Would there be an optimal "refinement" value of the cell size that would be a compromise between reliability and computing time? It actually depends on what we are looking for. As long as we don't bother whether or not we are heading towards a tipping point, there is apparently no problem. But if we were interested in the possible existence of a critical point at which the system would tip over, we would be facing a specific and interesting question. The approach of a critical point can be illustrated with the behaviour of falling dominoes. The fall off of one domino leads to that of several other ones, in some kind of avalanche. When we are far from the critical point (widely spaced dominos), avalanche sizes are small. For each of them, our computer system will be able to manage both the behaviour of each domino and its interactions with the other dominoes in the avalanche. 

But as we get closer to the critical point by tightening domino spacing, the avalanche sizes (number of dominoes or cells involved) increase, but the number of domino interactions "explodes". At the critical point, and by definition of this critical point, some avalanche sizes will tend to infinity (or at least reach the size of the system). The computing time for a system as vast and complex as the ocean-atmosphere ensemble will then tend towards a tremendous value, and at this stage the managers of the computer network would have already taken "simplification" measures to prevent system failure. It is therefore clear that the approach of a critical point as seen by such a numerical calculation looks like a horizon that vanishes as soon as one tries to reach it!

So, either there is no critical point in the immediate vicinity, or there is actually one. If there is none, numerical simulations are "reasonably" reliable. If there is one, they give at first sight a comparable result, but which might be misleading for the simple reason that they cannot predict "by construction" the existence of such a critical point. In this case, the computation result would be recurrently contradicted by events. This may possibly explain why forecasts have to be constantly revised upwards.

\section{The inverse method (critical phenomena)}
It is well known in Physics of Critical Phenomena that the approach of a critical point is announced by a series of fluctuations, whose amplitudes increase and eventually reach the system size when the tipping point is actually met, as mentioned above. This is the temporal equivalent of spatial fractal structures, that obey power-law statistical distributions with negative exponents. This phenomenon is called critical softening, due to the weakening of the restoring force in charge of bringing the system back to its initial equilibrium point, a force that vanishes at the critical transition. A simple explanation can be found in the attached video (\href{https://www.youtube.com/watch?v=OEHNb4yvkH4}{https://www.youtube.com/watch?v=OEHNb4yvkH4}).

The existence for the climate of such tipping point warning signals has already been considered since 2007 \citep{Livina&Lenton2007} and tested on \underline{paleoclimatic} events, which we know (in contrast with the present climate change) how they ended \citep{lenton2011,Lenton&al2012a,Lenton&al2012b,Lenton&al2019,Steffen&al2018}. Other works go in the same direction, and strongly suggest that the \underline{present} climate evolution, whose outcome is not yet known, could fit into this framework \citep{Bathiany&al2016, Louchet2016}. 

Are these perspectives really relevant to current climate evolution? We are indeed observing climate fluctuations of increasing amplitudes, in the form of extreme heat waves or cold snaps, unusual rainfalls or snowfalls, droughts, floods, hurricanes, forest fires, etc. It seems unlikely that a climate warming "limited" at the moment to 1$^o$C or 1.5$^o$C could be responsible for all these disturbances. Let us take the example of precipitations. For a typical temperature around 25$^o$C, with an atmospheric pressure $\rm{P_{atm}}$ of about 100 kPa, and using temperature variations of water vapour saturation pressure
, the increase of water vapour content in the vicinity of 25$^o$C can be easily estimated to no more than 5.10$^{-2}$ kPa /$^o$C, i.e. 5.10$^{-4}$ $\rm{P_{atm}}$ . The assertion (too frequently claimed) that a larger amount of water vapour contained in the atmosphere due to warming of a few degrees would significantly enhance the \underline{average} level of rain or snow precipitations is therefore unfounded. The average amount of precipitations should not be confused indeed with the succession (that results from critical softening) of periods of increasingly intense precipitations (that may be spectacular) separated by not less spectacular episodes of increasingly severe droughts.

The proximity of a critical point therefore seems more likely to account for these observations. We are probably facing an imminent climate bifurcation that will tip us into an unknown state. This assertion is confirmed by two recent papers \citep{neukom&al2019,st2019aberrant}. They highlight an "aberrant" and unprecedented global synchronization of the current climate warming, in contrast with the "little ice ages" of the 16th, 17th and 18th centuries, which occurred successively in different parts of the planet. In the same way as pre-critical fluctuations, this global synchrony is typical of the approach of a critical point for which the correlation length (i.e., the domino avalanche size) becomes of the order of magnitude of the system size.

A rather similar problem is that of the stability of glacier flow, which can also be described in terms of the approach of a critical point. The corresponding equations \citep{Faillettaz&al2015a,Faillettaz&al2016b} allow accurate predictions of icefall dates, and could be used for climate tipping, provided reliable instability data are available, as proposed in \cite{Louchet2016}. However, the obvious and impressive increase in climate fluctuations currently observed (in 2022) suggests that the tipping point is already here, which may explain again why every numerical prediction of temperature increase has to be revised upwards more and more frequently, giving increasingly alarming signals. 

In such a situation, we can predict with a fair degree of confidence that, when we will have switched into the next stable state, the pre-critical fluctuations (succession of heat waves, cold snaps, storms, etc...) will logically have disappeared (in contrast with what is usually believed), since we will be staying at the bottom of a new potential valley. The second quasi-certainty, which is in fact a truism, is that this state will be significantly warmer. However, since nobody has visited this unknown "territory" so far, any "extrapolation" of the present evolution is by nature doomed to failure. The temperature we would find there remains extremely difficult to predict. Nevertheless, it can be roughly estimated by comparison with the PETM (Paleocene-Eocene Thermal Maximum) episode, which occurred 56 million years ago \citep{McInerney2011}. At that time, a considerable volcanic activity yielded an injection of greenhouse gases into the atmosphere comparable to the present one (although for different reasons), but spread over 20,000 years. It resulted in an estimated global warming between 5$^o$C and 8$^o$C. The impressive abruptness of the present increase in greenhouse gas concentration (only 150 years), much shorter than "geological times", probably does not allow for a gradual adaptation of the system, and more particularly of the biosphere. This suggests a temperature response at least equal to the 5$^o$C to 8$^o$C of the PETM, and thus significantly beyond the 2$^o$C or 3$^o$C to which we are unsuccessfully "trying" to constrain the present evolution. More worryingly, present fluctuations suggest that such a situation is more likely to occur in the next few years rather than by the end of the century. This would be an irreparable and definitive disaster, that does not seem to be excluded if no drastic measures are decided and applied immediately at the planetary level, which is unfortunately doubtful. It should be noted that this is a global average, from which local climate developments could deviate, upwards or downwards.

This major tipping is already, and will be in the very near future, the consequence of a succession of elementary shifts of various sizes, each of them introducing an anticipated contribution to global irreversibility, and being as many positive feedback loops. One of them, currently in progress, reveals the major role that water can play in climate evolution. 

A remarkable feature of the Earth is indeed the presence of water in its three states, vapour, liquid and solid, with a proportion that depends on latitude and altitude. Water vapour is the dominant greenhouse gas in our atmosphere. This is why our planet is not a frozen desert. But, on the contrary, the reason why we are not living in a furnace is that water vapour and liquid water are in equilibrium, which has a self-stabilizing influence on its own greenhouse effect. Up to now indeed, due to the saturation of the atmosphere with water vapour in many parts of the world, any additional supply of water vapour at constant temperature is balanced on average by an equivalent amount of precipitation in the form of rain or snow (with the nuance that the level of water vapour saturation increases with temperature, which may contribute to a small but positive feedback loop, i.e. an autocatalytic effect in global warming).

And this is the point where, in addition to the balance between water vapour and liquid water, the balance between liquid water and ice also comes into play. Sea water is in contact with the pack ice in the Arctic Ocean, and with calving glaciers in Antarctica and Greenland, which keeps it locally around 0$^o$C. This cold water is then redistributed over the entire planet by the ocean stream network, slowing down ocean warming, and thus moderating the atmosphere saturation level in water vapour. However, the Arctic pack ice tends to disappear summer after summer. As for the retreat of the Antarctic and Greenland glaciers, the situation is increasingly alarming. The impressive Thwaites Glacier, that flows down into the Amundsen Sea, west of the Antarctic Peninsula, is a striking example. Its flow velocity continuously increases (this is probably another pre-critical phenomenon), and the "Thwaites Ice Sheet", which refers to its downstream and immersed part, threatens to break away from its upper part \citep{pettit2021,Alley&al2021,Wild&al2022}  
before 2030 and be dispersed as icebergs by ocean streams. As this immersed part currently holds back and slows down the upper part flow, the glacier will accelerate further after this breakup, its mass balance will collapse, and the glacier front will eventually stop being in contact (and thus in thermal equilibrium) with sea water. An additional zone of stabilization of the marine temperature will thus disappear, increasing further the warming rate of the ocean and the atmosphere. 

These prospects are all the more alarming as the current climate change is taking place over a dramatically short period of time. This situation may endanger humanity survival and absolutely requires an exceptional and immediate effort to reduce the concentration of greenhouse gases, along with a simultaneous, rapid and powerful take-off of resilience strategies.

\section{Summary and conclusion}
Current techniques for predicting climate change are mostly based on numerical modeling. However, the ocean-atmosphere system is what is usually referred to as a complex system, i.e. made up of a large number of interacting elements. These systems have been and still are thoroughly studied in theoretical physics. One of their well known properties is the "butterfly wing effect". In contrast with weather forecasting algorithms that may bypass to some extent the misleading consequences due to the butterfly effect over a few days or weeks, this is not the case for projections on a scale of years or decades. 

We have shown that in this case their sensitivity to initial conditions, as well as to all possible positive and negative feedbacks that are still to be discovered and introduced in algorithms, makes "massive" numerical modeling inoperative "by construction" for predicting climate change, because the management of cell interactions becomes out of control as soon as we approach the critical point.
Rather than trying to imagine on the basis of its past behaviour what Nature will do in the future, it seems more reasonable to listen to what it is already telling us about what it is doing now. Indeed, we should merely focus on climatic instabilities of all kinds and on their increasing intensities, announcing the approach of a critical point, and warning us that if we do not immediately take and apply drastic measures to reduce greenhouse gas emissions, we would readily and irreversibly tip over into a new and totally unknown state. Going beyond current estimates reported by IPCC, we can only say that temperatures in this state would be even higher than those predicted by the most recent (and probably future) updates of numerical simulations. 

Maintaining climate warming around 1.5$^o$C or 2$^o$C by 2030 or 2050 seems therefore to be \underline{fairly unrealistic in present conditions}. If no drastic measures are immediately taken, we would be heading to a warming of at least 5$^o$C or 8$^o$C in only a few years.

\bibliographystyle{copernicus}
\bibliography{bib_Louchet-Faillettaz}

%

\end{document}